\nonstopmode

\newcommand{\re}{{\rm Re\,}}
\newcommand{\im}{{\rm Im\,}}
\newcommand{\eV}{\U{eV}}
\newcommand{\rad}{\U{rad}}
\newcommand{\bohr}{\, a_0}
\newcommand{\Hartree}{\U{\mathit E_{\mathit h}}}
\newcommand{\degree}{{}^{\circ}}
\newcommand{\Cal}[1]{{\cal #1}}
\newcommand{\ie}{i.e.{}}
\newcommand{\eg}{e.g.{}}
\newcommand{\etal}{\textit{et al.}}
\newcommand{\U}[1]{\,{\rm{#1}}}
\newcommand{\I}[1]{_{\mathrm{#1}}}
\newcommand{\imag}{{\rm i}}
\newcommand{\euler}{\mathrm e}
\newcommand{\mat}[1]{\hbox{\boldmath{$#1$}\unboldmath}}
\newcommand{\Sum}{\sum\limits}
\newcommand{\Int}{\int\limits}
\newcommand{\Lim}{\lim\limits}
\newcommand{\transpose}{{}^{\textrm{\scriptsize T}}}
\newcommand{\unitop}{\hat{\mathbbm{1}}}
\newcommand{\unitmatrix}{\mat{\mathbbm{1}}}
\newcommand{\differential}{\>\mathrm d}
\newcommand{\bra}[1]{\left<\right.\!#1\!\left.\right|}
\newcommand{\ket}[1]{\left|\right.\!#1\!\left.\right>}
\newcommand{\bracket}[2]{\left<\right.\!#1\left.\right|#2\!\left.\right>}
\newcommand{\diag}{\textbf{diag}}
\newcommand{\kilobarns}{\U{kb}}
\newcommand{\Sxray}{$\sigma_{1s, \mathrm{n.l.}}(\omega_{\mathrm X})$}
\newcommand{\Spar} {$\sigma_{1s}^{\parallel}   (\omega_{\mathrm X})$}
\newcommand{\Sperp}{$\sigma_{1s}^{\perp}       (\omega_{\mathrm X})$}
\renewcommand{\frac}[2]{{{#1}\over{#2}}}

\documentclass[pra,aps,twocolumn,showpacs,nopreprintnumbers]{revtex4}
\usepackage{bm,bbm,amssymb,subeqnarray,graphicx}
\usepackage[nativepdf,bookmarks,bookmarksopen,bookmarksnumbered,raiselinks,%
pdftitle={Theory of x-ray absorption by laser-dressed atoms},%
pdfauthor={Christian Buth, Robin Santra},%
pdfsubject={Atomic physics},%
pdfkeywords={ab initio, laser dressing, x-rays, photoabsorption, cross section, %
polarization dependence, complex absorbing potentials, smooth exterior, %
complex scaling, non-Hermitian, perturbation theory, nonperturbative, %
laser-matter interaction, krypton}]{hyperref}

\begin{document}
\title{Theory of x-ray absorption by laser-dressed atoms}
\author{Christian Buth}
\thanks{Self-employed, Germany}
\author{Robin Santra}
\email[Corresponding author. Electronic address: ]{rsantra@anl.gov}
\affiliation{Argonne National Laboratory, Argonne, Illinois~60439, USA}
\date{09 April 2007}

\begin{abstract}
An \emph{ab initio} theory is devised for the x-ray photoabsorption cross section of atoms in
the field of a moderately intense optical laser ($800 \U{nm}$, $10^{13} \U{\frac{W}{cm^2}}$).
The laser dresses the core-excited atomic states, which introduces a dependence
of the cross section on the angle between the polarization vectors of the two
linearly polarized radiation sources.
We use the Hartree-Fock-Slater approximation to describe the atomic many-particle problem
in conjunction with a nonrelativistic quantum-electrodynamic approach
to treat the photon--electron interaction.
The continuum wave functions of ejected electrons are treated with a complex
absorbing potential that is derived from smooth exterior complex scaling.
The solution to the two-color (x-ray plus laser) problem is discussed in terms
of a direct diagonalization of the complex symmetric matrix representation of
the Hamiltonian.
Alternative treatments with time-independent and time-dependent non-Hermitian
perturbation theories are presented that exploit the weak interaction strength
between x~rays and atoms.
We apply the theory to study the photoabsorption cross section of krypton atoms near
the $K$~edge.
A pronounced modification of the cross section is found in the presence of the
optical laser.
\end{abstract}

%
%
%

\maketitle

\section{Introduction}

The ionization of an atom by a strong optical field may, under suitable conditions,
be described by a tunneling model~\cite{Delone:MP-00}.
The Ammosov-Delone-Krainov tunneling formula~\cite{Ammosov:TI-86}
predicts that ionization out of a sublevel with
an orbital angular momentum projection quantum number~$m = 0$ is strongly preferred
over ionization from~$m = \pm 1$~sublevels.
Employing a relatively intense laser, $I = 10^{14}$--$10^{15} \U{\frac{W}{cm^2}}$,
Young~\etal~\cite{Young:XR-06} studied laser-induced ionization of krypton
atoms from the $4p$~sublevel.
By monitoring the~$1s \to 4p$ resonance with a subsequent x-ray pulse at a photon
energy of~$14.3 \U{keV}$, they were able to measure a background-free signature
of the laser-produced $4p$~vacancy for several angles between laser and x-ray
polarizations.
The data exhibited a clear fingerprint of orbital alignment;
yet the x-ray absorption ratio between parallel and perpendicular polarizations
was significantly lower than that predicted by the nonrelativistic
tunneling picture~\cite{Ammosov:TI-86}.
By including the impact of spin-orbit coupling in the valence shell of krypton,
the experimental findings could be explained~\cite{Santra:SO-06}.

In the experiment of Young~\etal~\cite{Young:XR-06}, the laser was strong enough
to ionize the krypton atoms, so that x-ray absorption probed krypton ions.
A related scheme is the following.
If one overlaps the laser and x-ray fields in both space and time, but keeps
the laser intensity just low enough to avoid excitation of the closed-shell
atoms in their ground state, then the effect of the laser field is to modify
the final states that a core electron can reach via x-ray absorption.
This scenario---x-ray absorption by laser-dressed noble-gas atoms (see also
Ref.~\onlinecite{Buth:ET-up})---is the subject of an ongoing experiment
at Argonne National Laboratory and motivated our theoretical studies.

Certain aspects of the theory of x-ray absorption by laser-dressed atoms were
analyzed in Refs.~\onlinecite{Freund:TP-73,Ehlotzky:MC-75,Jain:XR-77,Ehlotzky:PE-78,%
Ferrante:EA-81,Ferrante:MP-81,Kalman:LA-88,Fonseca:XR-88,Kalman:XR-89,Kalman:HE-89,%
Cionga:TD-93,Glover:OL-96}.
Freund~\cite{Freund:TP-73} treats the simultaneous absorption of one laser photon
and one x-ray photon by solids.
The absorption of x~rays by laser-dressed hydrogen is examined in
Refs.~\onlinecite{Leone:TF-88,Kalman:HE-89,Cionga:TD-93}.
Particularly, Cionga~\etal~\cite{Cionga:TD-93} and K\'alm\'an~\cite{Kalman:HE-89}
point out the importance of laser-dressing effects close to the ionization
threshold.
Leone~\etal~\cite{Leone:TF-88} and Ferrante~\etal~\cite{Ferrante:EA-81} study the
angular distribution of the photoelectrons.
References~\onlinecite{Freund:TP-73,Ehlotzky:MC-75,Jain:XR-77,Ehlotzky:PE-78,%
Ferrante:EA-81,Ferrante:MP-81,Kalman:LA-88,Fonseca:XR-88,Kalman:XR-89,Kalman:HE-89,%
Cionga:TD-93,Glover:OL-96}
have in common that they treat the final state of the excited electron following
x-ray absorption essentially as a Volkov-type wave.
Some of them include a Coulomb correction.
They do not describe the element-specific properties of the x-ray absorption
cross section in the immediate vicinity of an inner-shell edge.

The energy spectrum of photoelectrons generated through XUV~photoionization of
helium in the presence of an intense laser field was measured in
Refs.~\onlinecite{Glover:OL-96,Guyetand:MC-05}.
The laser-induced modification of the x-ray absorption near-edge
structure~(XANES)~\cite{Rehr:TA-00,Als-Nielsen:EM-01} has not yet been
experimentally investigated for any laser-dressed atom or molecule.
In molecules, it must be expected that an external laser field will also have
an impact on the extended x-ray absorption fine
structure~(EXAFS)~\cite{Rehr:TA-00,Als-Nielsen:EM-01}.
Therefore, in addition to its fundamental interest, understanding the laser-dressing
effect on x-ray absorption is important from a practical point of view.
For instance, if one adiabatically aligns a molecule using an intense laser
pulse~\cite{Stapelfeldt:AM-03} and performs a XANES or EXAFS measurement in
order to determine molecular structure information, then one has to be able
to correct for the artificial impact of the aligning laser pulse on the
x-ray absorption cross section.

In this paper, we devise an \emph{ab initio} theory for the x-ray absorption
cross section of an isolated atom in the presence of an optical laser.
The Hartree-Fock-Slater mean-field model~\cite{Slater:AS-51,Slater:XA-72}
is utilized to treat the atomic many-electron problem.
This choice is adequate as shakeup and shakeoff effects are generally
weak in inner-shell photoionization.
They do not play a role in the immediate vicinity of the respective inner-shell edge.
To describe the radiation fields, we use a quantum-electrodynamic framework which
is equivalent to the semiclassical Floquet theory in the limit of high laser intensities.
The coupling of the x~rays to the atom is described perturbatively.
The laser dressing of the final-state manifold, however, is treated
nonperturbatively.
The theory is implemented in terms of the program \textsc{dreyd} as part of the
\textsc{fella} package~\cite{fella:pgm-06}.
We apply our method to study the x-ray absorption cross section of laser-dressed
krypton atoms near the $K$~edge.
Its dependence on the x-ray photon energy and on the angle between the
polarization vectors of the laser and the x~rays is investigated.

The article is structured as follows.
Section~\ref{sec:theory} discusses the theoretical foundation of the two-color
problem of an x-ray probe of a laser-dressed atom using an independent-particle
model for the atomic electrons, quantum electrodynamics for the photons, and
a complex absorbing potential for the continuum electron.
The conservation of the energy-integrated x-ray absorption cross section is also investigated.
Subsequently, the theory is applied to a krypton atom;
computational details are given in Sec.~\ref{sec:compdet};
the results are presented in Sec.~\ref{sec:results}.
Conclusions are drawn in Sec.~\ref{sec:conclusion}.

Our equations are formulated in atomic units.
The Bohr radius~$1 \U{bohr} = 1 \, a_0$ is the unit of length
and $1 \, t_0$ represents the unit of time.
The unit of energy is~$1 \U{hartree} = 1 \Hartree$.
Intensities are given in units of~$1 \Hartree \> t_0^{-1} \, a_0^{-2}
= 6.43641 \times 10^{15} \U{W \, cm^{-2}}$.

\section{Theory}
\label{sec:theory}
\subsection{Quantum electrodynamic treatment of atoms}

We solve the atomic many-electron problem in terms of a nonrelativistic
one-electron model.
Within this framework, each electron moves in the field of the atomic nucleus and
in a mean field generated by the other electrons.
The best such mean field derives from the Hartree-Fock method~\cite{Szabo:MQC-89}.
However, the Hartree-Fock mean field is nonlocal, due to the exchange interaction,
and therefore cumbersome to work with.
Slater~\cite{Slater:AS-51} introduced a local approximation to electron exchange,
which is the principle underlying the well-known $X\alpha$~method~\cite{Slater:XA-72}.
The resulting one-electron potential, $V_{\mathrm{HFS}}(r)$, is a central potential,
which satisfies
\begin{subeqnarray}
  V_{\mathrm{HFS}}(r) \rightarrow -\frac{Z}{r} & \quad \hbox{for} \quad & r \rightarrow 0 \; , \\
  \slabel{eq:V_HFS_far}
  V_{\mathrm{HFS}}(r) \rightarrow -\frac{1}{r} & \quad \hbox{for} \quad & r \rightarrow \infty
\end{subeqnarray}
for a neutral atom of nuclear charge~$Z$.
In this approximation, the atomic Hamiltonian is given by
\begin{equation}
  \label{eq:H_AT}
  \hat H_{\mathrm{AT}} = -\frac{1}{2} \vec\nabla^2 + V_{\mathrm{HFS}}(r) \; .
\end{equation}
In spherical polar coordinates, its eigenfunctions, the so-called atomic
orbitals, are the one-electron wave functions of the form~\cite{Merzbacher:QM-98}
\begin{equation}
  \label{eq:atomicsolution}
  \psi_{n,l,m}(r, \vartheta, \varphi) = \frac{u_{n,l}(r)}{r} \>
    Y_{l,m}(\vartheta, \varphi) \; .
\end{equation}
Here, $n$, $l$, and $m$ are the principal, orbital angular momentum, and
projection quantum number, respectively.
Using the ansatz~(\ref{eq:atomicsolution}) with the Hamiltonian~(\ref{eq:H_AT}), we
obtain the radial Schr\"odinger equation
\begin{equation}
  \label{eq:radial_H_AT}
  \Bigl[ -\frac{1}{2} \frac{\differential^2}{\differential r^2}
    + \frac{l(l+1)}{2r^2} + V_{\mathrm{HFS}}(r) \Bigr] \,
    u_{n,l}(r) = E_{n,l} \> u_{n,l}(r) \; ,
\end{equation}
where $E_{n,l}$~is the eigenenergy.
Equation~(\ref{eq:radial_H_AT}) is solved in a finite-element basis
set~\cite{Bathe:FE-76,Bathe:NM-76,Braun:FE-93,Ackermann:FE-96,%
Rescigno:CS-97,Meyer:TE-97,Santra:PM-04}---which is described in detail
in Ref.~\onlinecite{Santra:PM-04}---for~$l = 0, \ldots, n_l - 1$;
the positive integer~$n_l$ denotes the number of angular momenta included in
the basis set.
The calculated eigenfunctions satisfy the boundary conditions~$u_{n,l}
(r_{\mathrm{min}}) = 0$ and $u_{n,l}(r_{\mathrm{max}}) = 0$, where
$r_{\mathrm{min}} = 0$ and $r_{\mathrm{max}}$ is the maximum extension
of the radial grid.

Within the framework of quantum electrodynamics~\cite{Craig:MQ-84}, the Hamiltonian
describing the effective one-electron atom interacting with the electromagnetic field
reads
\begin{equation}
  \label{eq:Hamiltonian}
  \hat H_{\mathrm{QED}} = \hat H_{\mathrm{AT}} + \hat H_{\mathrm{EM}}
    + \hat H_{\mathrm I} \; .
\end{equation}
Here,
\begin{equation}
  \label{eq:H_EM}
  \hat H_{\mathrm{EM}} = \sum_{\vec k, \lambda} \omega^{\vphantom{\dag}}_{\vec k} \,
    \hat a^{\dag}_{\vec k, \lambda} \hat a^{\vphantom{\dag}}_{\vec k, \lambda}
\end{equation}
represents the free electromagnetic field;
its vacuum energy has been set to zero.
The operator~$\hat a^{\dag}_{\vec k, \lambda}$ ($\hat a^{\vphantom{\dag}}_{\vec k,
\lambda}$) creates (annihilates) a photon with wave vector~$\vec k$,
polarization~$\lambda$, and energy~$\omega_{\vec k} = c \, |\vec k| = |\vec k|
/ \alpha$ with the speed of light~$c$ and the fine-structure constant~$\alpha$.
The light-electron interaction term in electric-dipole approximation
is given in the length gauge by~\cite{Craig:MQ-84}
\begin{equation}
  \label{eq:eldipint}
  \hat H_{\mathrm I} = \vec x \cdot \sum_{\vec k, \lambda} \imag \,
    \sqrt{\frac{2\pi}{V} \, \omega^{\vphantom{\dag}}_{\vec k}} \>
    \bigl[ \vec e^{\vphantom{\ast}}_{\vec k, \lambda} \hat a^{\vphantom{\dag}}_{\vec k, \lambda}
         - \vec e^{\>\ast}_{\vec k, \lambda} \hat a^{\dag}_{\vec k, \lambda} \bigr] \; .
\end{equation}
We use the symbol~$\vec x = (x, y, z)\transpose$ for the atomic dipole operator
in Cartesian coordinates.
In Eq.~(\ref{eq:eldipint}), $V$~denotes the normalization volume of the
electromagnetic field and $\vec e_{\vec k, \lambda}$ indicates the
polarization vector of mode $\vec k,\lambda$.
Note that the electrons are treated in first quantization, whereas the electromagnetic
field is treated in second quantization.

The eigenstates of~$\hat H_{\mathrm{AT}} + \hat H_{\mathrm{EM}}$
may be written as a direct product of the form~$\ket{\psi_{n,l,m}} \,
\ket{\{N_{\vec k, \lambda}\}}$, where $\ket{\{N_{\vec k, \lambda}\}}$~is the
Fock state (or number state) of the photon field with $N_{\vec k, \lambda}$~photons
in the mode~$\vec k, \lambda$.
The curly braces indicate that more than one mode may be occupied.
The eigenfunctions of~$\hat H_{\mathrm{QED}}$ cannot, in general, be
written in the form~$\ket{\psi_{n,l,m}} \, \ket{\{N_{\vec k, \lambda}\}}$.
They may, however, be expanded in the basis~$\{\ket{\psi_{n,l,m}} \,
\ket{\{N_{\vec k, \lambda}\}}\}$, which we employ in the following.

\subsection{Complex absorbing potential}
\label{eq:CAP}

The absorption of photons may lead to the ejection of one or more electrons
from an atom;
either directly by photoionization or indirectly by the formation and decay of electronic
resonances.
The ejected electrons are in the continuum and thus their wave functions are
not square integrable~\cite{Kukulin:TR-89,Moiseyev:CS-98,Santra:NH-02}.
Therefore, they cannot be described by the basis set expansion techniques
in Hilbert space that are frequently employed in bound-state quantum
mechanics~\cite{Merzbacher:QM-98,Szabo:MQC-89}.
Several theories have been developed to make, particularly, resonance states,
nevertheless, amenable to a treatment with methods for bound states.
They typically lead to a non-Hermitian, complex-symmetric representation
of the Hamiltonian~\cite{Kukulin:TR-89,Moiseyev:CS-98,Santra:NH-02}.
In this framework, resonances are characterized by a complex energy
\begin{equation}
  \label{eq:siegert}
  E\I{res} = E\I{R} - \imag \, \Gamma / 2 \; ,
\end{equation}
which is frequently called Siegert energy~\cite{Kukulin:TR-89,Siegert:DF-39}.
Here, $\Gamma$ stands for the transition rate from the specific resonance
state to the continuum in which it is embedded.

Noteworthy for this work are complex scaling~\cite{Reinhardt:CC-82,%
Kukulin:TR-89,Moiseyev:CS-98} and complex absorbing
potentials~(CAP)~\cite{Goldberg:ML-78,Jolicard:OP-85,Jolicard:OP-86,%
Neuhauser:TD-89,Riss:CRE-93,Riss:RF-95,Moiseyev:DU-98,Riss:TC-98,Palao:CA-98,%
Palao:CC-98,Karlsson:AR-98,Sommerfeld:TA-98,Santra:EC-01,Santra:NH-02,%
Manolopoulos:DR-02,Poirier:SO-03,Poirier:SC-03}
which are exact methods to determine the resonance energies~(\ref{eq:siegert})
of a given Hamiltonian.
The CAPs have been analyzed thoroughly by Riss and Meyer~\cite{Riss:CRE-93} using
complex scaling.
Conversely, complex scaling of the Hamiltonian has been used to construct
a CAP that is adapted to a specific Hamiltonian~\cite{Moiseyev:DU-98,%
Riss:TC-98,Karlsson:AR-98}.
In all these methods, the resonance wave function associated with~$E\I{res}$,
Eq.~(\ref{eq:siegert}), is square-integrable.
To devise a CAP for~$\hat H_{\mathrm{QED}}$ in the spirit of
Refs.~\onlinecite{Moiseyev:DU-98,Riss:TC-98,Karlsson:AR-98}, we apply complex
scaling to it.
This is simply a complex coordinate transformation of the Hamiltonian.
Here, only the specialization to the scaling of the radial coordinate~$r = |\vec x|$
is needed, which proceeds in complete analogy to the one-dimensional case
of Refs.~\onlinecite{Moiseyev:DU-98,Karlsson:AR-98}.

The radial part~$r$ of the electron coordinates
is replaced by a path in the complex plane~$\varrho \equiv
F(r)$~\cite{Reinhardt:CC-82,Moiseyev:CS-98};
the resulting position vector is~$\vec \chi = \varrho \> (\cos \varphi \, \sin \vartheta, \, \sin \varphi \,
\sin \vartheta, \, \cos \vartheta)\transpose$ with the polar angle~$\vartheta$ and the
azimuth angle~$\varphi$~\cite{Arfken:MM-70}.
We use the path of Moiseyev~\cite{Moiseyev:DU-98} in the form of
Karlsson~\cite{Karlsson:AR-98}
\begin{equation}
  \label{eq:cmplxpath}
  F(r) = r + (\euler^{\imag\theta} - 1) \Bigl[r + r_0 + \frac{1}{2 \lambda}
    \ln \Bigl(\frac{1 + \euler^{2 \lambda (r - r_0)}}
              {1 + \euler^{2 \lambda (r + r_0)}}\Bigr) \Bigr] \; .
\end{equation}
Please refer to Refs.~\onlinecite{Moiseyev:DU-98,Karlsson:AR-98} for a graphical
representation.
The path starts at~$r = 0$ and runs along the positive real axis, \ie, $F(r) \approx r$.
In the vicinity of some distance~$r_0$ from the origin, the so-called exteriority,
it bends into the upper complex plane.
The bending is smooth, \ie,
$F(r)$~is infinitely many times continuously differentiable.
For~$r \gg r_0$, the path becomes the exterior scaling path, \ie, $F(r) \approx
r_0 + (r - r_0) \, \euler^{\imag\theta}$.
The parameter~$\lambda$ in Eq.~(\ref{eq:cmplxpath}) is a measure of how smooth
the bending around~$r_0$ is;
it is referred to as smoothness of the path.
A complex electron coordinate transformation of the Hamiltonian with a smooth path is
termed smooth exterior complex scaling~(SES)~\cite{Moiseyev:DU-98}.
Practical computational aspects of SES are discussed in Sec.~\ref{sec:compdet}.

Let us concentrate on the atomic contribution~$\hat H_{\mathrm{AT}}$ first.
The complex scaled radial Schr\"odinger equation is obtained by replacing~$r$
with~$\varrho$ in Eq.~(\ref{eq:radial_H_AT}).
It can be simplified following Karlsson~\cite{Karlsson:AR-98} (please note that there are
various misprints in the equations of Ref.~\onlinecite{Karlsson:AR-98}):
Letting $f(r) = F'(r)$ with ${}' = \frac{\differential}{\differential r}$, we make
the ansatz
\begin{equation}
  \label{eq:u_nl_mu_nl}
  u_{n,l}(\varrho) = \sqrt{f(r)} \; \frac{\mu_{n,l}(r)}{f(r)} \; .
\end{equation}
Applying the chain rule to rewrite the complex scaled Eq.~(\ref{eq:radial_H_AT})
with the substitution~(\ref{eq:u_nl_mu_nl}), we can extract expressions involving the
unscaled operator on the left-hand side of Eq.~(\ref{eq:radial_H_AT})
augmented by a CAP~\cite{Karlsson:AR-98}.
The CAP subsumes all corrective terms that arise from the complex scaled
kinetic energy.
A further contribution results from the atomic potential.
If the exteriority~$r_0$ is chosen sufficiently large, only the long-range
behavior of the atomic potential [cf.~Eq.~(\ref{eq:V_HFS_far})] is affected
by complex scaling.
Its contribution is added following Ref.~\onlinecite{Klaiman:RC-04}.
Finally, the CAP is given by
\begin{subeqnarray}
  \label{eq:SESCAP}
  \hat W &=& \hat W_{\rm k}
    + \Bigl(\frac{1}{2\varrho^2} - \frac{1}{2r^2}\Bigr) \, l \, (l+1)
    - \frac{1}{\varrho} + \frac{1}{r} \; , \\
  \slabel{eq:W_k}
  \hat W_{\rm k} &=& -\frac{1}{2} \frac{1}{f(r)} \frac{\differential^2}
    {\differential r^2} \frac{1}{f(r)} \\
  &&{} -\frac{1}{8} \frac{2 f''(r) f(r) - 3[f'(r)]^2}{f^4(r)}
   - \Bigl(-\frac{1}{2} \frac{\differential^2}{\differential r^2} \Bigr)
   \; . \nonumber
\end{subeqnarray}
In the interior, $r \ll r_0$, we have~$f(r) \approx 1$ and thus the scaled kinetic energy
becomes the unscaled one such that the correction term~$\hat W_{\rm k}$ vanishes.
Similarly, all other contributions to~$\hat W$ become negligible and $\hat W$ itself vanishes.
We will assume throughout that $r_0$~is large enough so that the occupied
atomic orbitals are unaffected by the~CAP.

The complex coordinate transformation of the radial Schr\"odinger
equation~(\ref{eq:radial_H_AT}) modifies the volume element in integrations involving
the~$u_{n,l}(\varrho)$;
it becomes~$f(r) \differential r$.
However, using the~$\mu_{n,l}(r)$ instead, the integration measure
becomes~$\differential r$.
Regarding the full Hamiltonian, $\hat H_{\mathrm{QED}}$, we note that the free
photon field, $\hat H_{\mathrm{EM}}$, does not depend on the electronic coordinates
and thus makes no contribution to~$\hat W$.
However, the interaction part, $\hat H_{\mathrm I}$, has to be complex scaled.
To keep the notation transparent, we refrain from formulating this transformation in terms
of a contribution to~$\hat W$ but apply complex scaling directly.

The CAP in Eq.~(\ref{eq:SESCAP}) is referred to as smooth exterior complex scaling
CAP~(SES-CAP).
It combines the advantages of simple polynomial CAPs~\cite{Riss:CRE-93}
on the one hand and complex scaling on the other hand,
eliminating many of their disadvantages.
First, no optimization with respect to a parameter is required
for SES-CAPs to determine resonance energies.
Second, the construction of a well-adapted CAP to a specific Hamiltonian is
rather straightforward.
Third, the resulting SES-CAP expressions are relatively simple and can be evaluated
efficiently on computers.

\subsection{X-ray probe of a laser-dressed atom}

In the following, only two modes (or two colors) of the radiation field are
considered:
The laser beam with photon energy~$\omega_{\mathrm L}$ and the x-ray beam
with photon energy~$\omega_{\mathrm X}$.
They are assumed to be monochromatic, linearly polarized, and copropagating.
The polarization vector $\vec e_{\mathrm{L}}$ of the laser defines the quantization
axis, which is chosen to coincide with the $z$~axis of the coordinate system.
Further, $\vec e_{\mathrm X}$ denotes the polarization vector of the x-ray beam
and $\vartheta_{\mathrm{LX}}$ is the angle between $\vec e_{\mathrm L}$ and
$\vec e_{\mathrm X}$, \ie, $\vec e_{\mathrm L} \cdot \vec e_{\mathrm X}
= \cos{\vartheta_{\mathrm{LX}}}$.
Let the photon numbers in the absence of interaction with the atom
be~$N_{\mathrm L}$ for the laser mode and $N_{\mathrm X}$ for the
x-ray mode, respectively.
The laser intensity is then given by
\begin{equation}
  I_{\mathrm L} = \frac{N_{\mathrm L}}{V}\frac{\omega_{\mathrm L}}{\alpha} \; ,
\end{equation}
with the fine structure constant~$\alpha = \frac{1}{c}$.
Similarly,
\begin{equation}
  \label{eq:J_X}
  J_{\mathrm X} = \frac{N_{\mathrm X}}{V} \frac{1}{\alpha}
\end{equation}
represents the x-ray photon flux.

As other modes do not contribute---radiative corrections are neglected---$\hat
H_{\mathrm{EM}}$ [Eq.~(\ref{eq:H_EM})] and the complex scaled $\hat
H_{\mathrm I}$ [Eq.~(\ref{eq:eldipint})] can be cast in a simplified form,
\begin{eqnarray}
  \hat H_{\mathrm{EM}} &=& \omega^{\vphantom{\dag}}_{\mathrm L} \,
    \hat a^{\dag}_{\mathrm L} \hat a^{\vphantom{\dag}}_{\mathrm L}
    + \omega^{\vphantom{\dag}}_{\mathrm X} \, \hat a^{\dag}_{\mathrm X}
    \hat a^{\vphantom{\dag}}_{\mathrm X} \; ,                      \\
  \hat H_{\mathrm I}  &=& \vec \chi\transpose \, \imag \, \sqrt{\frac{2\pi}{V} \,
    \omega^{\vphantom{\dag}}_{\mathrm L}}
    \, \bigl[ \vec e^{\vphantom{\ast}}_{\mathrm L} \hat a^{\vphantom{\dag}}_{\mathrm L}
            - \vec e^{\>\ast}_{\mathrm L}\hat a^{\dag}_{\mathrm L} \bigr] \nonumber \\
  &&{} + \vec \chi\transpose \, \imag \, \sqrt{\frac{2\pi}{V} \, \omega^{\vphantom{\dag}}_{\mathrm X}}
    \, \bigl[ \vec e^{\vphantom{\ast}}_{\mathrm X} \hat a^{\vphantom{\dag}}_{\mathrm X}
            - \vec e^{\>\ast}_{\mathrm X} \hat a^{\dag}_{\mathrm X} \bigr] \nonumber \\
  \label{eq:H1_HIL_HIX}
  &&{} = \hat H_{\mathrm {I, L}} + \hat H_{\mathrm {I, X}} \; .
\end{eqnarray}
Note that we rewrite the complex Hermitian scalar product in Eq.~(\ref{eq:eldipint})
in terms of a complex bilinear product here due to the complex
scaling~\cite{Kukulin:TR-89,Moiseyev:CS-98,Santra:NH-02}.
In comparison to all other interactions, the influence of the x-ray field may be
considered as weak.
We, therefore, separate the total complex scaled Hamiltonian~$\hat H_{\mathrm{QED}}$
[Eq.~(\ref{eq:Hamiltonian})] into a strongly interacting part,
\begin{equation}
  \label{eq:H_0}
  \hat H_0 = \hat H_{\mathrm{AT}} + \hat H_{\mathrm{EM}}
    + \hat H_{\mathrm {I, L}} + \hat W
\end{equation}
and a weakly interacting part
\begin{equation}
  \label{eq:H1_xrays}
  \hat H_1 = \hat H_{\mathrm {I, X}} \; .
\end{equation}
The SES-CAP~(\ref{eq:SESCAP}) contains the corrective terms that arise in the complex
scaling of~$\hat H_{\mathrm{AT}}$.
Note that $\hat H_0$ conserves the atomic angular momentum projection quantum number~$m$
and the number of x-ray photons~$N_{\mathrm X}$.
This partition of the Hamiltonian will prove useful below when perturbation
theory is applied to the problem.

We are concerned here with the case that $\omega_{\mathrm X}$ is large enough to
drive the excitation of an electron in the $K$~shell.
The x-ray intensity is assumed to be low enough to allow the description
of the interaction with the atom in terms of a one-photon absorption process.
This assumption is fully valid for experiments at third-generation synchrotron
radiation facilities,
but may have to be modified for
experiments with future free-electron lasers.
At such high photon energies, electrons in higher-lying shells are rather
insensitive to the x-ray field.
On the other hand, inner-shell electrons are unaffected by the laser.
As long as the laser intensity is small in comparison to an atomic unit, even the
valence shell is only weakly modified, and this modification is expected to be
similar before and after the absorption of an x-ray photon by a $K$-shell electron.

Hence, due to the weak coupling to the laser and the x~rays, we use a direct product
with the unperturbed $1s$~atomic orbital;
the initial state of the system before x-ray absorption reads
\begin{equation}
  \label{eq:Istate}
  \ket{I} = \ket{\psi_{1,0,0}} \ket{N_{\mathrm L}} \ket{N_{\mathrm X}} \; .
\end{equation}
It is an eigenvector of~$\hat H_{\mathrm{AT}} + \hat H_{\mathrm{EM}}$
with eigenvalue
\begin{equation}
 \label{eq:E_I}
  E_I = E_{1s} + N_{\mathrm L} \, \omega_{\mathrm L}
    + N_{\mathrm X} \, \omega_{\mathrm X} \; .
\end{equation}
It is also an approximate eigenvector of~$\hat H_0$ because the SES-CAP may be chosen
such that essentially it has no effect on~$\ket{I}$, \ie, $\bra{\psi_{1,0,0}} \hat W
\ket{\psi_{1,0,0}} \approx 0$ holds [see Sec.~\ref{eq:CAP}].
In Eq.~(\ref{eq:E_I}), $E_{1s}$~is the negative of the binding energy of a $K$-shell electron.
In principle, $E_{1s}$ is given by the energy of the atomic $1s$~orbital $E_{1,0}$.
Yet $E_{1,0}$ turns out to be not sufficiently accurate [see the caption
of Tab.~\ref{tab:RydTrans}].
To place the $K$~edge precisely, we replace~$E_{1,0}$
with the experimentally determined~$E_{1s}$.

In order to determine the manifold of laser-dressed final states, one needs to observe that
$N_{\mathrm X}$ is reduced by one unit after x-ray photon absorption and the final states are
assumed to be unperturbed by the x~rays.
Since~$\hat H_1$ couples only the electronic and x-ray degrees of freedom,
the accessible final states must have nonzero components with respect
to~$\ket{\psi_{n,l,m}} \ket{N_{\mathrm L}} \ket{N_{\mathrm X}-1}$, where~$l = 1$.
The projection quantum number~$m$ does not have to be zero,
for~$\vec e_{\mathrm X}$ does not necessarily coincide with~$\vec e_{\mathrm L}$, \ie,
the angle~$\vartheta_{\mathrm{LX}}$ does not have to be zero.
We employ the basis formed by the
\begin{equation}
 \label{eq:FloBasis}
  \ket{\Phi_{n,l,m,\mu}} = \ket{\psi_{n,l,m}} \ket{N_{\mathrm L} - \mu}
    \ket{N_{\mathrm X} - 1} \; ,
\end{equation}
where the quantum numbers~$n$, $l$, and $m$ correspond to orbitals that
are unoccupied in the atomic ground state.
The number of laser photons that are absorbed (emitted) by the core-excited
electron is denoted by~$\mu = 0, \pm 1, \pm 2, \ldots \;$.
The operator~$\hat H_{\mathrm{AT}} + \hat H_{\mathrm{EM}}$ is diagonal in this
basis with eigenvalues~$E_{n,l,\mu} = E_{n,l} + (N_{\mathrm L}-\mu)
\, \omega_{\mathrm L} + (N_{\mathrm X}-1) \, \omega_{\mathrm X}$;
the operator~$\hat H_0$, however, is not.
A global energy shift
\begin{equation}
  \hat H_{\mathrm{EM}}' = \hat H_{\mathrm{EM}} - N_{\mathrm L} \omega_{\mathrm L}
    - [N_{\mathrm X}-1] \omega_{\mathrm X}
\end{equation}
makes the notation more transparent.
It carries over---using a definition analogous to Eq.~(\ref{eq:H_0})---to~$\hat H_0$,
which becomes~$\hat H_0'$.
Thus
\begin{subeqnarray}
  \relax
  \label{eq:H_0prime}
  \hat H_0' \ket{I} &=& E_I' \ket{I} \; , \\
  \slabel{eq:E_Ip}
  E_I' &=& E_{1s} + \omega_{\mathrm X} \; , \\
  \relax
  [\hat H_{\mathrm{AT}} + \hat H_{\mathrm{EM}}'] \ket{\Phi_{n,l,m,\mu}} &=&
    [E_{n,l} - \mu \, \omega_{\mathrm L}] \ket{\Phi_{n,l,m,\mu}} \; .
\end{subeqnarray}

The only nonvanishing matrix elements of~$\hat H_0'$ with respect to the basis
$\left\{\ket{\Phi_{n,l,m,\mu}}\right\}$ are
\begin{widetext}
  \begin{subeqnarray}
    \label{eq:H_0_Floquet}
    \bra{\Phi_{n,l,m,\mu}} \hat H_0' \ket{\Phi_{n',l,m,\mu}}
      &=& [E_{n,l} - \mu \, \omega_{\mathrm L}] \, \delta_{n,n'}
      + \bra{\psi_{n,l,m}} \hat W \ket{\psi_{n',l,m}} \; , \\
    \slabel{eq:FloCoup}
    \bra{\Phi_{n,l,m,\mu}} \hat H_0' \ket{\Phi_{n',l',m,\mu\pm 1}}
      &=& \sqrt{2 \pi \alpha I_{\mathrm L}} \, \bra{\psi_{n,l,m}}
      \varrho \, \cos \vartheta \ket{\psi_{n',l',m}} \; .
  \end{subeqnarray}
\end{widetext}
It has been exploited in the coupling matrix elements~(\ref{eq:FloCoup}) that
the laser is linearly polarized along the $z$~axis of the coordinate system, \ie,
in terms of spherical polar coordinates~$\varrho \, \cos \vartheta
= \vec \chi\,\transpose \, \vec e_{\mathrm L}$ holds.
Moreover, the number of photons in the laser mode is assumed to be much
greater than one.
Note that~$\hat H_{\mathrm {I, L}}$ [Eq.~(\ref{eq:H1_HIL_HIX})] in~$\hat H_0'$
produces an extra factor~$\mp\imag$ which is not present in Eq.~(\ref{eq:FloCoup}).
To remove this factor, we observe that Eq.~(\ref{eq:H_0_Floquet}) forms a block-tridiagonal
matrix with respect to the photon number~$\mu$.
The rows and columns of the block matrices are labeled by the orbital quantum
numbers~$n,l,m$ and $n',l',m$, respectively.
Let $\mat U = \diag(\unitmatrix, \imag \, \unitmatrix, \imag^2 \, \unitmatrix, \ldots,
\imag^{n\I{ph}} \, \unitmatrix)$ be a unitary transformation,
with the number of photon blocks being~$n\I{ph}$.
The unit matrices~$\unitmatrix$ have the dimension of the number of atomic
orbitals~(\ref{eq:atomicsolution}) used.
Applying $\mat U$ to the original matrix with additional $\mp\imag$~factors, here
denoted by~$\mat F$, yields the matrix without them, $\mat F'$ [Eq.~(\ref{eq:H_0_Floquet})],
\ie, $\mat U^{\dagger} \mat F \mat U = \mat F'$.
The matrix representation~$\mat F'$ of~$\hat H_0'$ is of the
Floquet type~\cite{Shirley:SE-65,Chu:IF-77,Chu:TS-85,Burke:RM-91,Dorr:RM-92,Chu:BF-04}.
See, for example, Refs.~\onlinecite{Kulander:MI-87,Kulander:TD-88,Huens:ND-97,%
Taylor:MA-99,Kamta:MS-02} and references therein for other computational
approaches to atomic strong-field physics.
Furthermore, the matrix representation~(\ref{eq:H_0_Floquet}) is block-diagonal with
respect to the projection quantum number~$m$ because $m$ is a conserved quantity for
linearly polarized light.
Hence it is sufficient to focus on the subblocks
\begin{equation}
  (\mat H_0^{\prime(m)})_{(n,l,\mu),(n',l',\mu')} =
    \bra{\Phi_{n,l,m,\mu}} \hat H_0' \ket{\Phi_{n',l',m,\mu'}} \; ,
\end{equation}
for each~$m$.
They are evidently rather sparse.
The rows and columns of~$\mat H_0^{\prime(m)}$ are labeled by the triple
index~$(n, l, \mu)$.

All $K$-shell-excited states undergo rapid relaxation via Auger decay or x-ray emission;
in the latter case primarily by $K\alpha$~fluorescence.
As these relaxation pathways are many-particle phenomena, they are not included in our
one-particle description.
To take these effects into consideration, we note that the decay of a $K$-shell hole involves
primarily other inner-shell electrons;
the excited electron is a spectator.
It is, therefore, reasonable to assign a width~$\Gamma_{1s}$ to each excited one-particle
level associated with a core hole in the many-particle wave function.
In a very good approximation, $\Gamma_{1s}$ may be assumed to be independent
of the laser field and the quantum numbers of the spectator electron.
We replace $\mat H_0^{\prime(m)}$ by
\begin{equation}
  \label{eq:H0_Gamma}
  \mat H_0^{(m)} = \mat H_0^{\prime(m)} - \imag \, \frac{\Gamma_{1s}}{2} \,
    \unitmatrix \; .
\end{equation}
If the original~$\mat H_0^{\prime(m)}$ is diagonalizable~%
\footnote{Unlike a real-symmetric matrix, a complex symmetric matrix is not
necessarily diagonalizable~\cite{Santra:NH-02}.
Yet we assume this property throughout and verify it in practical
computations.},
so is~$\mat H_0^{(m)}$.
Given the generally complex eigenvalues of~$\mat H_0^{\prime(m)}$, the
energies~$E_F^{\prime(m)}$, the eigenvalues of~$\mat H_0^{(m)}$ are simply~$E_F^{(m)}
= E_F^{\prime(m)} - \imag \, \frac{\Gamma_{1s}}{2}$.
The eigenvectors~$\vec c_F^{\>(m)}$ satisfy
\begin{equation}
  \label{eq:diag_H0}
  \mat H_0^{(m)} \vec c_F^{\>(m)} = E_F^{(m)} \vec c_F^{\>(m)} \; .
\end{equation}
They are normalized and form a complex orthogonal set~$\vec c_F^{\>(m)}\transpose \,
\vec c_{F'}^{\,(m)} = \delta_{F,F'}$~\cite{Santra:NH-02}.
The
vector~$\vec c_F^{\>(m)}$ defines a laser-dressed state with respect to the
basis~(\ref{eq:FloBasis}),
\begin{equation}
  \label{eq:dressedket}
  \ket{F^{(m)}} = \sum_{n,l,\mu} c_{n,l,\mu,F}^{(m)}
    \ket{\psi_{n,l,m}} \ket{N_{\mathrm L}-\mu} \ket{N_{\mathrm X}-1} \; .
\end{equation}
In view of the complex orthogonality of the eigenvectors of ${\bm H}_0^{(m)}$,
the bra vector associated with~$\ket{F^{(m)}}$ is
\begin{equation}
  \label{eq:dressedbra}
  \bra{F^{(m)}} = \sum_{n,l,\mu} c_{n,l,\mu,F}^{(m)}
  \bra{\psi_{n,l,m}} \bra{N_{\mathrm L} - \mu} \bra{N_{\mathrm X}-1} \; ,
\end{equation}
\ie, the coefficients~$c_{n,l,\mu,F}^{(m)}$ are left complex-unconjugated.
With this definition, it follows that~$\bracket{F^{(m)}}{F'^{(m')}}
= \delta_{F,F'} \, \delta_{m,m'}$.

Having determined the relevant eigenstates of the x-ray unperturbed
Hamiltonian~$\hat H'_0$, \ie, the final states reached by x-ray absorption from the
ground state, we are now in the position to explore the effect of~$\hat H_1$
[Eq.~(\ref{eq:H1_xrays})].
Let $\mat H_0 = \diag \bigl( \ldots, \mat H^{(-1)}_0, \mat H^{(0)}_0, \mat H^{(1)}_0,
\ldots \bigr)$ be the matrix representation of the
unperturbed Hamiltonian constructed from Eqs.~(\ref{eq:H_0_Floquet}) and
(\ref{eq:H0_Gamma}).
In principle, one can proceed in complete analogy to the previous paragraphs,
by augmenting the matrix~$\mat H_0$ with the additional matrix elements involving the
initial state [Eqs.~(\ref{eq:Istate}) and (\ref{eq:E_Ip})]
\begin{subeqnarray}
  \bra{I} \hat H'_0 \ket{I} &=& E_I' \; , \\
  \slabel{eq:xflocoup}
  \bra{\Phi_{n,l,m,\mu}} \hat H_1 \ket{I}
    &=& \delta_{\mu,0} \sqrt{2 \pi \alpha \, \omega_{\mathrm X} J_{\mathrm X}}
    \nonumber \\
  &&{} \times \bra{\psi_{n,l,m}} \vec \chi\transpose \, \vec e_{\mathrm X}
    \ket{\psi_{1,0,0}} \; .
\end{subeqnarray}
A unitary transformation was applied as in Eq.~(\ref{eq:FloCoup}) to remove the
$\mp\imag$~factors in Eq.~(\ref{eq:xflocoup}).
We obtain the matrix representation~$\mat H$ of the full Hamiltonian, including all
energy shifts, in the basis~$\{\ket{I}, \ket{\Phi_{n,l,m,\mu}}\}$
\begin{equation}
  \label{eq:fullhammat}
  \mat H = \left(
  \begin{array}{cc}
    E_I' & \mat H\transpose_{0 \, I} \\
    \mat H_{0 \, I} & \mat H_0
  \end{array}
  \right) \; ,
\end{equation}
with~$(\mat H_{0 \, I})_{n,l,m,\mu} = \bra{\Phi_{n,l,m,\mu}} \hat H_1 \ket{I}$.
Diagonalizing~$\mat H$ and examining its eigenvectors, one determines the
eigenvalue~$E_I''$ that corresponds to the eigenvector with the largest overlap
with~$\ket{I}$.
The eigenvalue~$E_I''$ is a Siegert energy~(\ref{eq:siegert});
the imaginary part, $\im E_I'' = -\frac{\Gamma_I}{2}$,
yields the transition rate from~$\ket{I}$ to any of the accessible final states.
It allows one to obtain the x-ray photoabsorption cross section via
\begin{equation}
  \label{eq:xsect}
  \sigma_{1s} = n_{1s} \, \frac{\Gamma_I}{J_{\mathrm X}} \; .
\end{equation}
The additional factor, $n_{1s} = 2$, accounts for the number of electrons in the
$K$~shell because the $1s$~atomic orbital is used to form the initial state~$\ket{I}$.

The matrix~$\mat H$ represents the most general formulation of the interaction of
two-color light with atoms.
It can easily be generalized to study multiphoton x-ray physics by allowing for the
absorption and emission of several x-ray photons in the basis~(\ref{eq:FloBasis}).
Although straightforward, the (partial) diagonalization of~$\mat H$ is quite costly.
Additionally, we are interested in the dependence of the cross section on the x-ray energy,
which requires a sampling of~$\omega_{\mathrm X}$ for a range of values.
Above all, $\mat H$ does not immediately reveal the underlying physics, \ie, the dependence on
the angle between the polarization vectors of the x-ray beam and the laser
beam [Sec.~\ref{sec:pacs}] as well as the approximate conservation of the
integrated cross section [Sec.~\ref{sec:cics}].

These aspects can be addressed by a perturbative treatment of the x-ray--electron interaction
pursued in the ensuing Secs.~\ref{sec:rspt} and \ref{sec:tdpt}.
We give a time-independent and a time-dependent derivation.
The first route is logically simpler but we anticipate the reasoning to
be less well known than the reasoning in the second route which is easier to
understand intuitively.
However, because of the non-Hermiticity involved, the second route requires special care.
To treat the absorption of an x-ray photon with perturbation theory, the
Hamiltonian is represented in the eigenbasis of the unperturbed part~$\mat H_0$, \ie,
$\{\ket{I}, \ket{F^{(m)}}\}$ [Eqs.~(\ref{eq:Istate}), (\ref{eq:dressedket}), and
(\ref{eq:dressedbra})].
A single reference perturbation theory is sufficient because the diagonalization
of~$\mat H_0$ already incorporates the strong laser-atom interaction.

\subsection{Time-independent treatment}
\label{sec:rspt}

The time-independent, non-Hermitian Rayleigh-Schr\"odinger perturbation
theory of Ref.~\onlinecite{Buth:NH-04} is applied to study the x-ray absorption.
Up to second order, the effect of~$\hat H_1$ on the energy of the single
initial state~$\ket{I}$, Eq.~(\ref{eq:Istate}), is given by
\begin{subeqnarray}
  \slabel{eq:srpt-E0}
  E_{I,0} &=& \bra{I} \hat H'_0 \ket{I} = E'_I \; , \\
  \slabel{eq:srpt-E1}
  E_{I,1} &=& \bra{I} \hat H_1 \ket{I} = 0 \; , \\
  \slabel{eq:srpt-E2}
  E_{I,2} &=& \sum_{m, F} \frac{\bra{I} \hat H_1 \ket{F^{(m)}}
    \bra{F^{(m)}} \hat H_1 \ket{I}} {E_{I,0} - E_F^{(m)}} \; .
\end{subeqnarray}
The first order correction~(\ref{eq:srpt-E1}) vanishes due to
the fact that the matrix representation of the perturbation~$\hat H_1$ in
Eq.~(\ref{eq:H1_xrays}) has vanishing diagonal elements.
This is because $\hat H_1$ consists of a linear combination of an x-ray
photon creation operator and an annihilation operator.
The transition rate~$\Gamma_I$ from~$\ket{I}$ to any other state
results from the imaginary part of the Siegert energy~(\ref{eq:siegert}):
\begin{equation}
  \label{eq:G_I_indep}
  \begin{array}{rcl}
    \displaystyle \Gamma_I &=& \displaystyle -2 \  \im [E_{I,0}
      + E_{I, 1} + E_{I, 2}] \\
    &=& \displaystyle 2 \  \im \Sum_{m, F} \frac{\displaystyle \bra{I} \hat H_1
      \ket{F^{(m)}} \bra{F^{(m)}} \hat H_1 \ket{I}} {\displaystyle E_F^{(m)}
      - E'_I} \; .
  \end{array}
\end{equation}
Note that the unperturbed energy~$E'_I$ in Eq.~(\ref{eq:E_Ip}) is real.

\subsection{Time-dependent treatment}
\label{sec:tdpt}

Alternatively, the x-ray photoabsorption rate can be derived by
judicious application of time-dependent perturbation
theory~\cite{Sakurai:MQM-94} or the closely related method of the variation
of constants of Dirac~\cite{Craig:MQ-84} to approximate solutions to
the time-dependent Schr\"odinger equation.
Here, we pursue the latter route.
At~$t = 0$, the system is in state~$\ket{I}$.
A general state ket (or wave packet) is given by
\begin{equation}
  \label{eq:wavepacket}
  \ket{\Psi,t} = \beta_I(t) \> \euler^{-\imag E'_I t} \ket{I}
  + \sum_{m,F} \beta_F^{(m)}(t) \> \euler^{-\imag E_F^{(m)} t} \ket{F^{(m)}} \; ,
\end{equation}
where $\{\ket{I}, \ket{F^{(m)}}\}$ forms an orthonormal eigenbasis of~$\hat H''_0
= \hat H'_0 - \imag \, \frac{\Gamma_{1s}}{2} \, (\unitop - \ket{I} \bra{I})$
[see Eqs.~(\ref{eq:H_0prime}) and (\ref{eq:H0_Gamma})].
Inserting formula~(\ref{eq:wavepacket}) into the time-dependent Schr\"odinger
equation~$(\hat H''_0 + \hat H_1) \ket{\Psi,t} = \imag \frac{\partial}{\partial t}
\ket{\Psi,t}$ and exploiting~$\hat H''_0 \ket{n}
= E_n \ket{n}$ for~$\ket{n} \in \{\ket{I}, \ket{F^{(m)}}\}$, we arrive at the
equation of motion for the expansion coefficients~$\beta_{n}(t)$
by projecting on the~$\bra{n}$:
\begin{equation}
  \label{eq:eqomo}
  \imag \, \dot \beta_n(t) = {\mathrm{e}}^{\imag E_n t} \bra{n} \hat H_1 \ket{\Psi,t} \; .
\end{equation}
The matrix element in this expression can be rewritten immediately in terms of
the basis kets~$\{\ket{I}, \ket{F^{(m)}}\}$ by inserting Eq.~(\ref{eq:wavepacket}).
The resulting equations are integrated analytically for all~$F,m$, employing the initial
conditions~$\beta_I(0) = 1$ and $\beta_F^{(m)}(0) = 0$
to obtain first order corrections for the coefficients~$\{\beta_F^{(m)}(t)\}$.
In the non-Hermitian case considered here, the textbook
strategy~\cite{Sakurai:MQM-94,Merzbacher:QM-98} of using~$\Lim_{t \to \infty}
\beta_F^{(m)}(t)$ to construct the transition amplitude cannot be applied:
Because $\im E_F^{(m)} < 0$ due to Eq.~(\ref{eq:siegert}) and $\Gamma > 0$, the
amplitude~$\beta_F^{(m)}(t)$~diverges in the limit~$t \to \infty$.
This causes no difficulty, for the physically relevant quantity is the ground-state
amplitude~$\beta_I(t)$, more precisely~$\dot{\beta}_I(t)$.
By inserting the coefficients~$\beta_F^{(m)}(t)$ and Eq.~(\ref{eq:wavepacket}) into
Eq.~(\ref{eq:eqomo}) with~$n = I$, and exploiting Eq.~(\ref{eq:srpt-E1}),
the equation of motion of~$\beta_I(t)$, to second order in the
perturbation~$\hat H_1$, is found to be
\begin{equation}
  \dot{\beta}_I(t) = \imag \Sum_{m,F}
  \frac{\bra{I} \hat H_1 \ket{F^{(m)}} \bra{F^{(m)}} \hat H_1 \ket{I}}
       {E_F^{(m)} - E'_I}
  \bigl[ 1 - \euler^{\imag (E'_I - E_F^{(m)}) t} \bigr] \; .
\end{equation}

The probability of finding the atom in the initial state is
\begin{equation}
  P_I(t) = \beta_I^{\ast}(t) \, \beta_I(t) \; .
\end{equation}
Consequently, the negative of the x-ray absorption rate is
\begin{eqnarray}
  \dot{P}_I(t) &=& \dot{\beta}_I^{\ast}(t) \, \beta_I(t)
    + \beta_I^{\ast}(t) \, \dot{\beta}_I(t) \nonumber \\
  & \approx & \dot{\beta}_I^{\ast}(t) + \dot{\beta}_I(t) \\
  & = & 2 \  \re \dot{\beta}_I(t) \; . \nonumber
\end{eqnarray}
Here, the center line follows from the weakness of x-ray absorption,
\ie, $\beta_I(t) \approx {\mathrm{const}}$ for all~$t$.
At~$t = 0$, the absorption rate vanishes, \ie, $\dot{\beta}_I(0) = 0$.
For $t \gg \frac{1}{\Gamma_{1s}}$, the coefficient $\dot{\beta}_I(t)$---and
hence the absorption rate---becomes stationary.
In this limit,
\begin{equation}
  \label{eq:G_I_td}
  -\Gamma_I = \dot{P}_I = -2 \  \im \biggl[ \sum_{m,F}
    \frac{\bra{I} \hat H_1 \ket{F^{(m)}} \bra{F^{(m)}} \hat H_1
    \ket{I}}{E_F^{(m)} - E'_I} \biggr] \; ,
\end{equation}
which is equivalent to Eq.~(\ref{eq:G_I_indep}).

\subsection{Photoabsorption cross section}
\label{sec:pacs}

Combining Eqs.~(\ref{eq:J_X}), (\ref{eq:H1_xrays}), (\ref{eq:Istate}), (\ref{eq:E_Ip}),
(\ref{eq:dressedket}), and (\ref{eq:dressedbra}) with Eq.~(\ref{eq:G_I_indep})---or
equivalently, with Eq.~(\ref{eq:G_I_td})---the $1s$~absorption cross section
is obtained using Eq.~(\ref{eq:xsect}):
\begin{equation}
  \label{eq:sigcomb}
  \sigma_{1s} =
    4 \pi \, n_{1s} \, \alpha \, \omega_{\mathrm X} \  \im \biggl[ \sum_{m,F}
    \frac{(\mathcal{D}_F^{(m)})^2}{E_F^{(m)} - E_{1s} - \omega_{\mathrm X}}
    \biggr] \; ,
\end{equation}
where
\begin{equation}
  \label{eq:reddip}
  \mathcal{D}_F^{(m)} = \Sum_n c_{n,1,0,F}^{(m)} \bra{\psi_{n,1,m}}
    \vec \chi\transpose \, \vec e_{\mathrm X} \ket{\psi_{1,0,0}}
\end{equation}
is a complex scaled transition dipole matrix element between the $1s$~one-particle
state in the atomic ground state and the $F$th laser-dressed atomic state
with projection quantum number~$m$.
An expression that is formally similar to Eq.~(\ref{eq:sigcomb})
has been obtained by Rescigno~\etal~\cite{Rescigno:RM-75,Rescigno:CS-76}
for the photoabsorption cross section without laser-dressing
using a semiclassical treatment of the radiation field and
time-dependent perturbation theory.
The extra factor~$n_{1s}$ does not appear in Refs.~\onlinecite{Rescigno:RM-75,Rescigno:CS-76}
because, there, the equations are formulated using many-particle wave functions.
Instead, we use expressions for orbitals and, hence, have to sum over the two equal
contributions from both $K$-shell electrons.

The radial part of the integrals in Eq.~(\ref{eq:reddip}) is given by
\begin{equation}
  \label{eq:radialdip}
  \mathcal{R}_n = \int_0^{\infty} u_{n,1}(r) \, \varrho \, u_{1,0}(r) \differential r \; .
\end{equation}
Although we express~$\mathcal{R}_n$ in terms of the complex path~(\ref{eq:cmplxpath}),
the actual result does not noticeably depend on it.
The compactness of the $1s$~atomic orbital restricts the integrand in Eq.~(\ref{eq:radialdip})
to a region near the nucleus where~$\varrho \approx r$ [see Sec.~\ref{eq:CAP}].
The angular part of the dipole matrix elements in Eq.~(\ref{eq:reddip}) between~$s$ and
$p$~spherical harmonics is found to be~$\frac{1}{3} \varkappa_m(\vartheta_{\mathrm{LX}})$ with
\begin{equation}
  \label{eq:pacangdep}
  \varkappa_m(\vartheta_{\mathrm{LX}}) = \left\{
  \begin{array}{ll}
    \frac{1}{2} \sin^2(\vartheta_{\mathrm{LX}}) & , \; m = +1 \; , \\
    \hphantom{\frac{1}{2}} \cos^2(\vartheta_{\mathrm{LX}}) & , \; m = \hphantom{+} 0 \; , \\
    \frac{1}{2} \sin^2(\vartheta_{\mathrm{LX}}) & , \; m = -1 \; .
  \end{array}
  \right.
\end{equation}
Thus the x-ray absorption cross section is finally
\begin{equation}
  \label{eq:phxsect}
  \begin{array}{rcl}
    \displaystyle \sigma_{1s}(\omega_{\mathrm{X}}, \vartheta_{\mathrm{LX}})
      &=& \displaystyle \frac{4 \pi}{3} \, n_{1s} \, \alpha \, \omega_{\mathrm X}
      \Sum_{m=-1}^1 \varkappa_m(\vartheta_{\mathrm{LX}}) \; \\
    &&{} \displaystyle \times \im \Bigl[\Sum_F
      \frac{(d_F^{(m)})^2}{E_F^{(m)} - E_{1s} - \omega_{\mathrm X}} \Bigr] \; ,
  \end{array}
%
\end{equation}
where
\begin{equation}
  d_F^{(m)} = \sum_n c_{n,1,0,F}^{(m)} \, \mathcal{R}_n \; .
\end{equation}
This expression explicitly spells out the dependence of~$\sigma_{1s}$ on the
angle between the laser and x-ray polarizations.
Notice that the summands in Eq.~(\ref{eq:phxsect}) for~$m = 1$ and $m = -1$ are
equal due to the cylindrical symmetry of the problem.
We can simplify Eq.~(\ref{eq:phxsect}) further;
with the definitions
\begin{subeqnarray}
  \sigma_{1s}^{\parallel}(\omega_{\mathrm X}) &\equiv&
    \sigma_{1s}(\omega_{\mathrm X},  0 \degree) \; , \\
  \sigma_{1s}^{\perp}    (\omega_{\mathrm X}) &\equiv&
    \sigma_{1s}(\omega_{\mathrm X}, 90 \degree) \; ,
\end{subeqnarray}
we obtain the simple expression
\begin{equation}
  \label{eq:phax}
  \sigma_{1s}(\omega_{\mathrm X}, \vartheta_{\mathrm{LX}})
    = \sigma_{1s}^{\parallel}(\omega_{\mathrm X}) \cos^2\vartheta_{\mathrm{LX}}
    + \sigma_{1s}^{\perp}    (\omega_{\mathrm X}) \sin^2\vartheta_{\mathrm{LX}} \; .
\end{equation}
For vanishing laser intensity, we have $\sigma_{1s}^{\parallel}(\omega_{\mathrm X})
= \sigma_{1s}^{\perp}(\omega_{\mathrm X})$ and thus the angular dependence disappears,
\ie, the cross section becomes a circle in a polar plot with
radius~$\sigma_{1s}^{\parallel}(\omega_{\mathrm X})$.
Generally, Eq.~(\ref{eq:phax}) describes  an ellipse in a polar plot.

The origin of the difference between the cross sections~$\sigma_{1s}^{\parallel}
(\omega_{\mathrm X})$ and $\sigma_{1s}^{\perp}(\omega_{\mathrm X})$ in the presence
of a laser field can be understood in terms of the structure of the Floquet
matrix~$\mat F'$ in Eq.~(\ref{eq:H_0_Floquet});
it is block-diagonal with respect to the projection quantum number~$m$.
Clearly, only the block with~$m = 0$ contains $s$~states.
Therefore, the $m = 0$~block is distinguished from all other blocks of~$\mat F'$.
For parallel laser and x-rays polarizations, $m$~of the total
system~(\ref{eq:fullhammat}) is a conserved quantum number.
Hence excitations out of the $1s$~initial state into the final state manifold,
spanned by the eigenstates of~$\mat F'$, couple exclusively to the $m = 0$~block,
which is reflected by the factor~$\varkappa_m(0\degree) = \delta_{m,0}$
in Eq.~(\ref{eq:phxsect}).
In the case of perpendicular polarization vectors, $m$~is no longer conserved;
only the final states from the blocks of~$\mat F'$ with~$m = \pm 1$ contribute
because~~$\varkappa_m(90\degree) = \frac{\delta_{|m|,1}}{2}$.
The different structure of the blocks of~$\mat F'$ leads to different
matrix elements and thus different final states.

The form~(\ref{eq:phax}) of the angular dependence of the total cross section
is obtained in electric dipole approximation for the radiation--electron interaction
in the coupling Hamiltonian~(\ref{eq:eldipint}).
Electron correlations and nondipole effects, primarily for the x~rays~\cite{Krassig:ND-03},
can be expected to lead to a deviation from this formula.
As it is easier to measure a total cross section than it is to determine an
angular resolved photoelectron distribution, \eg, Ref.~\onlinecite{Krassig:ND-03},
laser dressing opens up another route to study such effects.

\subsection{Conservation of the integrated cross section}
\label{sec:cics}

Let us investigate under which approximations the integrated
photoabsorption cross section, \ie,
\begin{equation}
  \label{eq:intcross}
  \begin{array}{rcl}
    \displaystyle S &=& \displaystyle\Int_0^{\infty} \sigma_{1s}(\omega_{\mathrm X})
      \differential \omega_{\mathrm X} \\
    &=& \displaystyle 2 \, \frac{n_{1s}}{J_{\mathrm X}} \; \im \Biggl[ \Sum_{m,F}
      \Int_0^{\infty} \frac{\bra{I} \hat H_1 \ket{F^{(m)}} \bra{F^{(m)}}
      \hat H_1 \ket{I}}{E_F^{(m)} - E_{1s} - \omega_{\mathrm X}} \differential
      \omega_{\mathrm X} \Biggr] \; ,
  \end{array}
\end{equation}
is independent of the intensity of the dressing laser,
where we use Eqs.~(\ref{eq:E_Ip}) and (\ref{eq:xsect}) in conjunction with
Eq.~(\ref{eq:G_I_indep}) or, equivalently, with Eq.~(\ref{eq:G_I_td}).
The integral in Eq.~(\ref{eq:intcross}) is known to converge because
at photon energies much higher than the $K$~edge, the cross section is well known to decay
rapidly~\cite{Bethe:QM-57} [see also Eq.~(\ref{eq:xsect_decay}) and
the surrounding discussion].

Most of the contributions to~$S$ arise in the vicinity of the $K$~edge because there the product
of transition matrix elements, $\bra{I} \hat H_1 \ket{F^{(m)}} \bra{F^{(m)}} \hat H_1
\ket{I}$, is large due to compact Rydberg states and low-energy continuum states.
The product contains a factor~$\omega_{\mathrm X}$ from the
$\hat H_1$~operators~(\ref{eq:H1_xrays}).
This dependence on the x-ray photon energy is eliminated by replacing the factor
with~$-E_{1s}$, \ie, let $\hat H'_1 \equiv \frac{\hat H_1}{\sqrt{\omega_{\mathrm X}}}$,
then $\hat H_1 \approx \sqrt{-E_{1s}} \> \hat H'_1$ holds.
With this approximation, the integral over the photoabsorption cross section is
\begin{equation}
  \label{eq:sepintcr}
  S = -2 E_{1s} {n_{1s} \over J_{\mathrm X}} \> \im \Bigl[ \Sum_{m,F} \bra{I}
    \hat H'_1 \ket{F^{(m)}} \bra{F^{(m)}} \hat H'_1 \ket{I} I_F^{(m)} \Bigr] \, ,
\end{equation}
with
\begin{equation}
  \label{eq:I_Fm}
  I_F^{(m)} = \Int_{-R}^R \frac{1}{E_F^{(m)} - E_{1s} - \omega_{\mathrm X}}
    \differential \omega_{\mathrm X} \; .
\end{equation}
Here, we extend the integration range to negative values, which does not change~$I_F^{(m)}$
noticeably because the real part of the pole position is much larger than zero.
Moreover, we refrain from taking the limit~$R \to \infty$ to avoid
divergences in intermediate expressions.

The integral~(\ref{eq:I_Fm}) is rewritten by closing the contour in the lower
complex $\omega_{\mathrm X}$~plane in a semicircle.
Let $\Cal C$ be the full contour and $\cup$ be the semicircle;
then we have~$I_F^{(m)} = I_{F, \Cal C}^{(m)} - I_{F, \cup}^{(m)}$.
The integral~$I_{F, \Cal C}^{(m)}$ is evaluated easily with the residue
theorem~\cite{Arfken:MM-70}, yielding~$I_{F, \Cal C}^{(m)} = 2 \pi \imag$,
where an extra negative sign comes from the clockwise integration along~$\Cal C$.
The contour integral over the semicircle, \ie, $\omega_{\mathrm X} =
R \, \euler^{-\imag \phi}$, is
\begin{equation}
  I_{F, \cup}^{(m)} = \Int_0^{\pi} \frac{-\imag}{R^{-1} \, \euler^{\imag \phi} \,
    (E_F^{(m)} - E_{1s}) - 1} \differential \phi \; .
\end{equation}
Letting $R$ become much larger than all of the~$|E_F^{(m)} - E_{1s}|$, the integral
becomes~$I_{F, \cup}^{(m)} = \imag \, \pi$.
With this we obtain $I_F^{(m)} = \imag \, \pi$, which is independent
of~$m$ and $F$.
Hence, $I_F^{(m)}$ can be eliminated from the sum in Eq.~(\ref{eq:sepintcr}).

The sum over products of matrix elements in Eq.~(\ref{eq:sepintcr}) can be expressed as
\begin{equation}
  \label{eq:Sdef}
  \begin{array}{rl}
     \Sum_{m,F} \bra{I} \hat H'_1 \ket{F^{(m)}}
      \bra{F^{(m)}} \hat H'_1 \ket{I} & \\
     {} + \bra{I} \hat H'_1 \ket{I}
      \bra{I} \hat H'_1 \ket{I}
      = & \bra{I} \hat H'_1 \, \hat{\Cal P}  \, \hat H'_1 \ket{I}
  \end{array}
\end{equation}
by adding the term~$\bra{I} \hat H'_1 \ket{I}^2 = 0$
[cf.~Eq.~(\ref{eq:srpt-E1})].
Equation~(\ref{eq:Sdef}) is rewritten in terms of the projector, $\hat{\Cal P}$,
which projects on the subspace that is spanned by the basis~$\{\ket{I},
\ket{F^{(m)}}\}$.
Here, $\hat{\Cal P}$ can be formulated in terms of the basis
functions~(\ref{eq:Istate}) and (\ref{eq:FloBasis})
\begin{equation}
  \hat{\Cal P} = \sum_{n, l, m, \mu} \ket{\Phi_{n,l,m,\mu}} \bra{\Phi_{n,l,m,\mu}}
    + \ket{I} \bra{I}
\end{equation}
using Eqs.~(\ref{eq:dressedket}) and (\ref{eq:dressedbra}).
Inserting the definition of~$\hat H'_1$ into Eq.~(\ref{eq:Sdef}), we arrive with
Eq.~(\ref{eq:J_X}) [cf.~also Eq.~(\ref{eq:xflocoup})] at
\begin{equation}
  \label{eq:Sform}
  \bra{I} \hat H'_1 \, \hat{\Cal P} \, \hat H'_1 \ket{I}
    = 4 \pi \, \alpha \, J_{\mathrm X} \, \bra{\psi_{1,0,0}}(\vec \chi\transpose
    \, \vec e_{\mathrm X})^2
    \ket{\psi_{1,0,0}} \; ,
\end{equation}
exploiting that $\vec e_{\mathrm X}$~denotes a real vector.

Gathering the results in Eqs.~(\ref{eq:sepintcr}), (\ref{eq:Sdef}), and
(\ref{eq:Sform}), we find
\begin{equation}
  \label{eq:SdepS}
  S = -8 \pi^2 \, n_{1s} \, \alpha \, E_{1s} \; \re \bra{\psi_{1,0,0}} (\vec \chi\transpose \,
    \vec e_{\mathrm X})^2
    \ket{\psi_{1,0,0}} \; .
\end{equation}
Clearly, the integrated photoabsorption cross section does not depend on the intensity
of the dressing laser or the angle between polarization vectors of x~rays and laser.
Therefore, within the approximations made, the integrated cross section is conserved.

\section{Computational details}
\label{sec:compdet}

\begin{figure}
  \begin{center}
    \includegraphics[clip,width=\hsize]{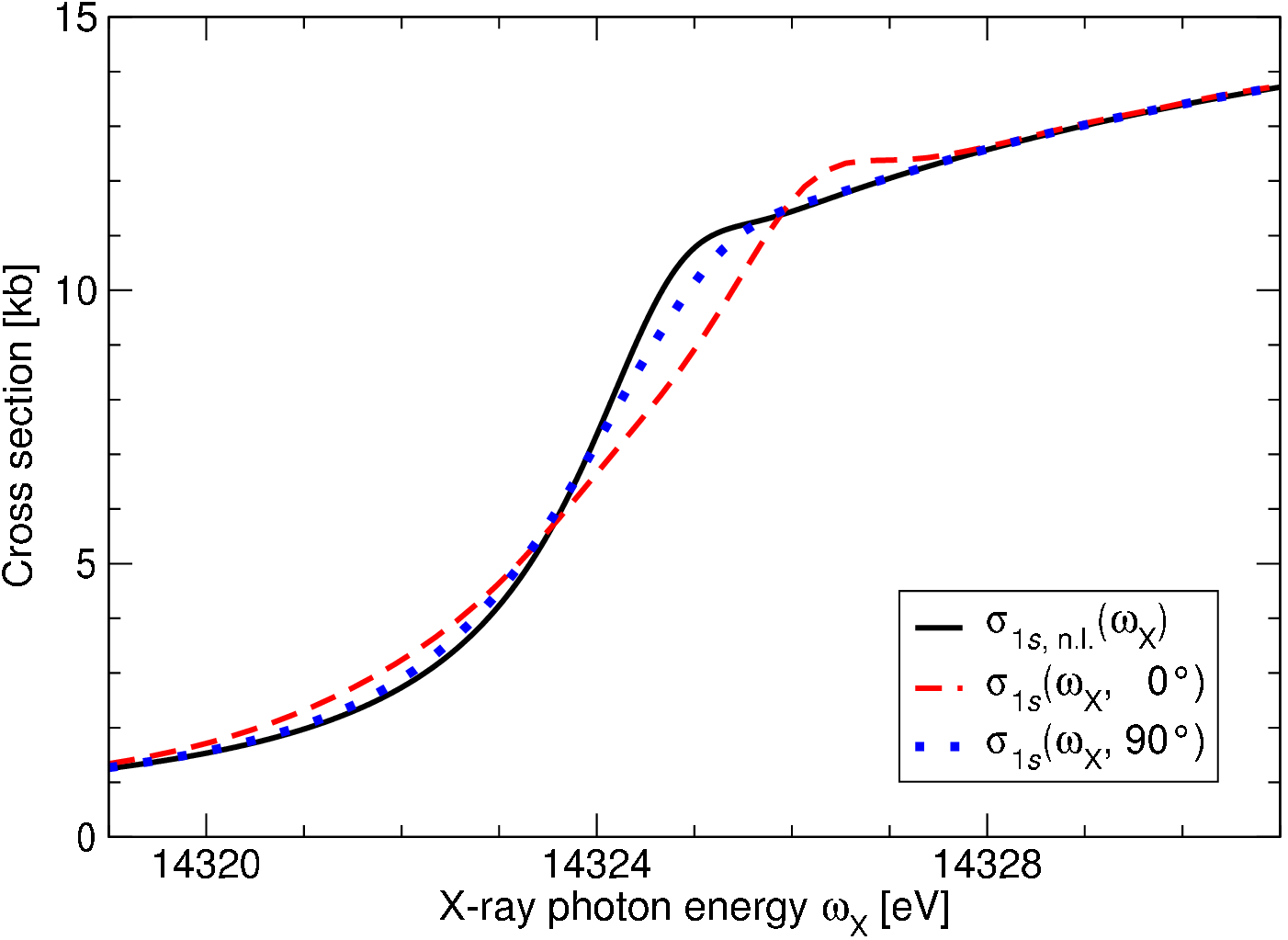}
    \caption{(Color online) X-ray photoabsorption cross section of the krypton atom near the
             $K$~edge with laser dressing~$\sigma_{1s}(\omega_{\mathrm X},
             \vartheta_{\mathrm{LX}})$ and without it~\Sxray{}.
             The angle~$\vartheta_{\mathrm{LX}}$ is formed between the polarization
             vectors of the laser and the x~rays.
             The laser operates at a wavelength of~$800 \U{nm}$
             with an intensity of~$10^{13} \U{W \over cm^2}$.}
    \label{fig:xsection}
  \end{center}
\end{figure}

The theory of the previous Sec.~\ref{sec:theory} shall now be applied to study
the interaction of a krypton atom with two-color light.
We use the Hartree-Fock-Slater code written by Herman and Skillman~\cite{Herman:AS-63},
which has proven advantageous for atomic photoionization studies, \eg,
Ref.~\onlinecite{Manson:PI-68}, to determine the one-particle
potential~$V_{\mathrm{HFS}}(r)$ of krypton in Eq.~(\ref{eq:H_AT}).
As in the original program of Herman and Skillman, the $X\alpha$ parameter is
set to unity, in accordance with Ref.~\onlinecite{Slater:AS-51}.
The radial equation~(\ref{eq:radial_H_AT}) is solved using a representation
of~$u_{n,l}(r)$ in terms of $3001$~finite-element functions, which span a radial grid
from~$r_{\mathrm{min}} = 0$ to $r_{\mathrm{max}} = 60 \bohr$.
For each of the orbital angular momentum quantum numbers~$l = 0, 1, 2, 3$
considered, the lowest $100$~solutions were computed and used to form atomic
orbitals~(\ref{eq:atomicsolution}) in the following;
except in Fig.~\ref{fig:xsect_decay}, where we use $500$~solutions to reproduce
the high-energy behavior.
In all cases, we verified that our results are converged with respect to the atomic
basis set.

\begin{figure}
  \begin{center}
    \includegraphics[clip,width=\hsize]{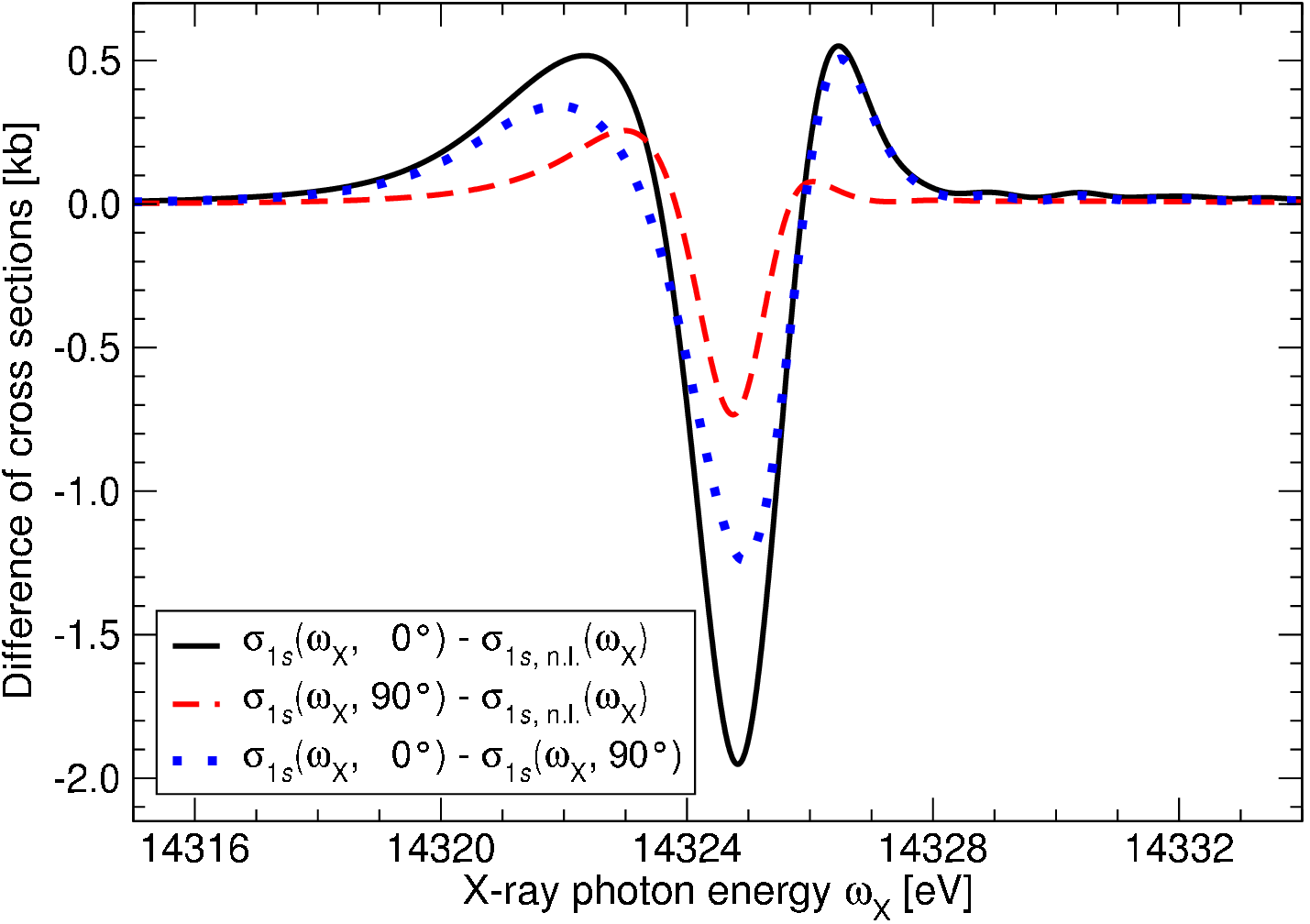}
    \caption{(Color online) Difference between x-ray photoabsorption cross sections
         of the krypton atom near the $K$~edge.
         Symbols as in Fig.~\ref{fig:xsection}.}
    \label{fig:diffsigma}
  \end{center}
\end{figure}

The SES-CAP is constructed using the complex path~(\ref{eq:cmplxpath}).
The path requires care when evaluated numerically due to the exponential
functions therein.
A complex scaling angle of~$\theta = 0.13 \rad$ is used and
the smoothness is~$\lambda = 5 \bohr^{-1}$.
In our computations, the Hartree-Fock-Slater atomic potential
assumes the long-range limit~(\ref{eq:V_HFS_far})
to eight significant digits after~$3 \bohr$, which defines the
inner region of the krypton atom.
We choose the exteriority~$r_0 = 7 \bohr$, which ensures that the atomic ground state
is unperturbed by the SES-CAP, as exploited in the derivation of Eq.~(\ref{eq:SESCAP}).

The laser is assumed to operate with an optical wavelength of~$800 \U{nm}$
(photon energy~$1.55 \eV$) and an intensity of~$10^{13} \U{W \over cm^2}$.
The x-ray photon energy is varied in the vicinity of the $K$~edge of krypton,
for which we use the experimental value~$E_{1s} = -14327.17 \U{eV}$ of
Breinig~\etal~\cite{Breinig:AI-80}.
The experimental value for the decay width of a hole in the krypton
$K$~shell is~$\Gamma_{1s} = 2.7$~eV~\cite{Krause:NW-79,Chen:RK-80}.
To describe the laser dressing accurately, we have to include the photon blocks
with~$\mu = 0, \pm 1, \ldots, \pm 5$ in the Floquet matrix~(\ref{eq:H_0_Floquet}).

\section{Results and discussion}
\label{sec:results}

The x-ray photoabsorption cross section of krypton is plotted in
Fig.~\ref{fig:xsection} for three different cases:
(a)~The cross section without laser dressing, \Sxray,
(b)~the cross section for parallel polarization vectors, denoted
by~\Spar{}${} = \sigma_{1s}(\omega_{\mathrm X}, 0 \degree)$, and
(c)~the cross section for perpendicular polarization vectors, denoted
by~\Sperp{}${} = \sigma_{1s}(\omega_{\mathrm X}, 90 \degree)$.
Following Eq.~(\ref{eq:phax}), the chosen angles exhibit the largest effect of
the polarization dependence of the cross section of the laser-dressed atom.
The impact of the laser dressing is most clearly reflected in
differences of the photoabsorption cross sections.
They are shown in Fig.~\ref{fig:diffsigma} for the
cases~\Spar${}-{}$\Sxray, \Sperp${}-{}$\Sxray, and \Spar${}-{}$\Sperp.

Inspecting Fig.~\ref{fig:xsection}, we see that, beginning a few electronvolts
below the $K$~edge, the cross sections for laser on and off are smaller
than~$2 \kilobarns$.
The reason for this is the fact that one is away from resonances for such x-ray energies
because no one-photon excitation and ionization processes out of Kr$\,1s$~states
are energetically allowed.
Excitations or ionizations of higher lying shells, $L$, $M$, \ldots, do not contribute
noticeably in the energy range shown and are, therefore, not included in our theory.

\begin{table}
  \centering
  \begin{ruledtabular}
    \begin{tabular}{lcc}
      Transition & $E_{\mathrm{HFS}}$ [eV] & $E_{\mathrm{expt}}$ [eV] \\
      \hline
      $1s \rightarrow 5p$ & 14324.81 & 14324.57 \\
      $1s \rightarrow 6p$ & 14326.01 & 14325.86 \\
      $1s \rightarrow 7p$ & 14326.48 & 14326.45 \\
      $1s \rightarrow 8p$ & 14326.75 & 14326.72
    \end{tabular}
  \end{ruledtabular}
  \caption{Transition energies from the $K$~shell of krypton to Rydberg orbitals.
           Our results, $E_{\mathrm{HFS}}$, are obtained using the
           Hartree-Fock-Slater energies of the Rydberg orbitals~$E_{\mathrm{Ryd}}$
           in terms of the formula~$E_{\mathrm{HFS}} = E_{\mathrm{Ryd}} - E_{1s}$,
           where $E_{1s}$~is the $K$-shell energy.
           The Hartree-Fock-Slater value for~$E_{1s}$ is~$E_{1,0} = -14022.88 \U{eV}$;
           it is replaced by the precise experimental value~$-14327.17
           \U{eV}$~\cite{Breinig:AI-80}.
           The experimental values~$E_{\mathrm{expt}}$ are taken from Tab.~2
           in Breinig~\etal~\cite{Breinig:AI-80}.}
  \label{tab:RydTrans}
\end{table}

In the vicinity of the $K$~edge, there is an appreciable impact of the
laser dressing on~\Spar{} [Figs.~\ref{fig:xsection} and \ref{fig:diffsigma}];
it is suppressed with respect to the laser-free curve~\Sxray{}
between~$\approx 14323 \eV$ and $\approx 14326 \eV$.
Outside of this range, the cross section~\Spar{} is somewhat larger than~\Sxray.
Here, \Sperp{} behaves in a similar way, yet with a significantly
lower deformation of the curve in relation to~\Sxray.

To understand this behavior we need to investigate the electronic structure
in the vicinity of the ionization threshold first.
Close to the threshold but still below are the energies for the
transitions to Rydberg states.
Exemplary Rydberg transition energies are listed in Tab.~\ref{tab:RydTrans}.
Comparing the theoretical values to the experimental values, we
find that the Hartree-Fock-Slater approximation describes the
Rydberg orbitals, $5p$, $6p$, $7p$, and $8p$, accurately.
This is attributed to the property of such orbitals to be very
extended with only a small amplitude in the vicinity of the nucleus.
Consequently, the one-particle approximation is very well justified.
This reasoning is supported additionally by the observation that the
agreement between theoretical and experimental energies in
Tab.~\ref{tab:RydTrans} increases with increasing principal
quantum number of the Rydberg orbital involved.

Inspecting Tab.~\ref{tab:RydTrans}, we notice that the dip at~$14324.82 \eV$
($14324.72 \eV$) for the solid black (dashed red) curve lies very close to the energy of the
$1s \to 5p$~Rydberg transition, \ie, to the energy of the final state which is
a $1s^{-1} \, 5p$~configuration.
In a lowest order perturbation theoretical argument, emission of a laser photon from
the $1s^{-1} \, 5p$~configuration leads to the energy~$14323.27 \eV$ ($14323.37 \eV$).
It agrees with the energy of $1s^{-1} \, 5s$~configuration, $14323.67 \eV$.
Conversely, the absorption of a laser photon from the $1s^{-1} \, 5p$~configuration leads
to~$14326.37 \eV$ ($14326.27 \eV$), which is in the range of the energies of the
$1s^{-1} \, 4d$ and $1s^{-1} \, 5d$~configurations, at $14325.61 \eV$ and
$14326.29 \eV$, respectively.
However, the coupling matrix elements between~$5p$ and $5d$ and higher
$d$~orbitals are small compared with the coupling of $5p$ and $4d$~orbitals.
Hence, we conclude that the laser dressing causes a strong coupling of the
$1s^{-1} \, 5p$~configuration to the $1s^{-1} \, 5s$ and
$1s^{-1} \, 4d$~configurations,
which leads to the suppression of the $1s \to 5p$~transition and an enhancement around the
energies of the $1s^{-1} \, 5s$ and $1s^{-1} \, 4d$~configurations.

Further above the $K$~edge, we see in Fig.~\ref{fig:xsection} that the cross sections for
laser on and off are essentially the same.
Obviously, the relative importance of the energetic shift of the continuum of final
states due to the laser dressing, the ponderomotive potential~\cite{Freeman:PP-88}
$U\I{p} = 2 \pi \alpha \frac{I\I{L}}{\omega^2\I{L}} = 0.60 \eV$,
decreases for increasing x-ray photon energies.
This is quantified by the quotient of~$U\I{p}$ and the energy of the ejected electron.
Clearly, the latter energy grows with increasing x-ray energy.
In Fig.~\ref{fig:diffsigma}, above~$14327.17 \eV$, weak wiggles with the spacing
of roughly the laser photon energy of~$1.55 \eV$ are observed.

%
%
%
%
%
%
%
%
%
%
%
%
%
%
%
%
%
%
%
%
%
%
%
%
%
%
%
%
%
%
The conservation of the integrated cross section~(\ref{eq:intcross}) is proven
under certain approximations in Sec.~\ref{sec:cics}.
The applicability of this theorem can be examined by a numerical integration
of the curves in Fig.~\ref{fig:diffsigma} for $\omega_{\mathrm X}$ in the
range~$14300 \eV$ to $14400 \eV$.
We obtain~$-0.197 \kilobarns \, \eV$ (solid black), $-0.062 \kilobarns \, \eV$
(dashed red), and $-0.135 \kilobarns \, \eV$ (dotted blue).
Following Sec.~\ref{sec:cics}, the resulting value---for an integration from~$0$
to $\infty$---should be zero.
To put these values in relation with the total deviation of the curves from zero,
an integration of the absolute value of the curves is performed which
yields~$5.119 \kilobarns \, \eV$ (solid black), $1.534 \kilobarns \, \eV$ (dashed
red), and $3.679 \kilobarns \, \eV$ (dotted blue).
We find that the proportion of the integrated cross section to the integrated
absolute value of the cross section is less than~$-0.002$ for all curves.
Hence the assumptions made in Sec.~\ref{sec:cics} are fulfilled
very well and the integrated cross section is essentially independent of the
dressing-laser intensity.

\begin{figure}
  \begin{center}
    \includegraphics[clip,width=\hsize]{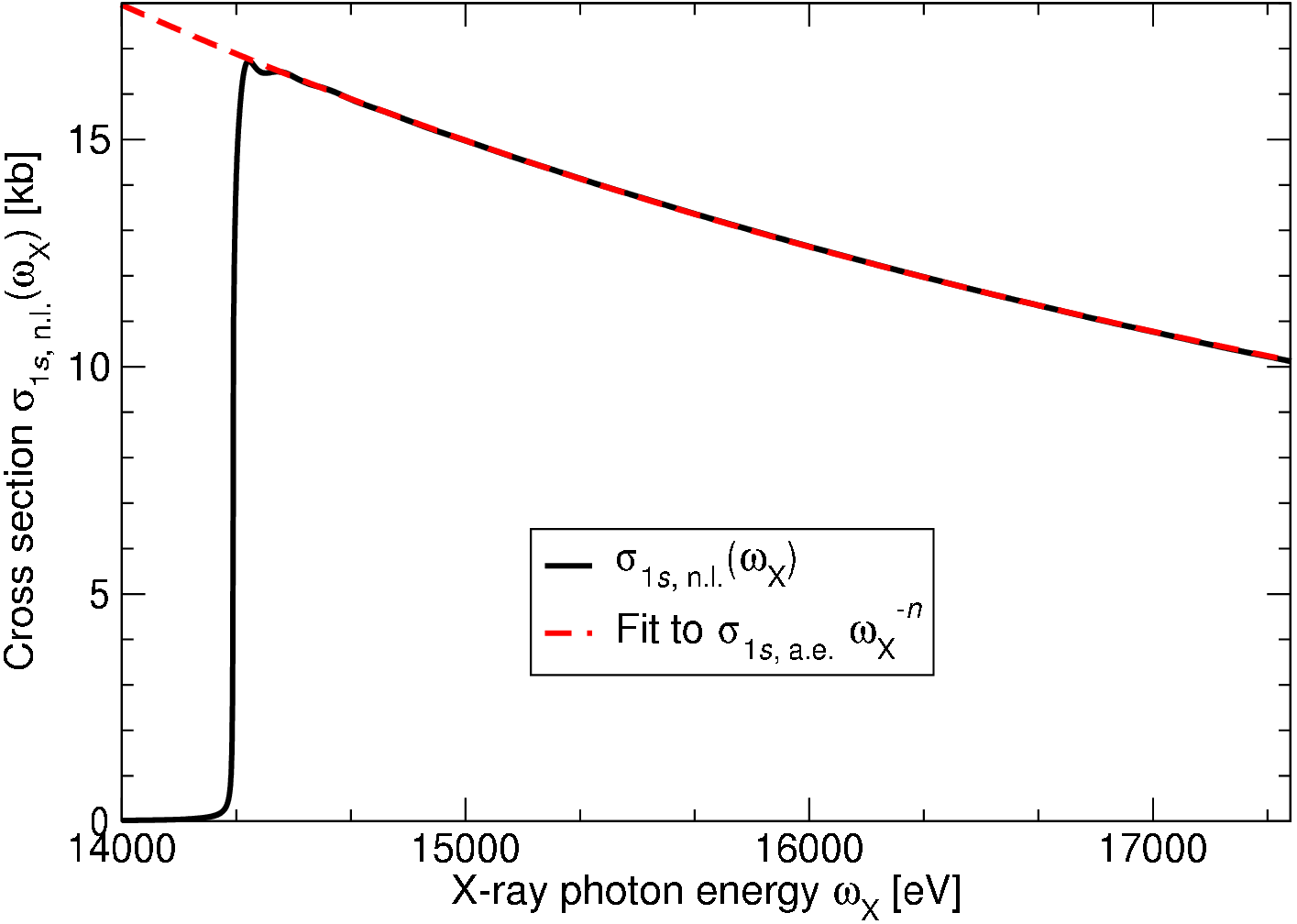}
    \caption{(Color online) Behavior of the photoabsorption cross section without
         laser~\Sxray{} of the krypton atom above the $K$~edge.
         The data are fitted to the ansatz in Eq.~(\ref{eq:xsect_decay}),
         which gives~$\sigma_{1s,\; \mathrm{a.e.}} = 18623.8 \kilobarns \, \eV^n$
         and $n = 2.63$.}
    \label{fig:xsect_decay}
  \end{center}
\end{figure}

The photoabsorption cross section without laser~\Sxray{} above the $K$~edge is
displayed in Fig.~\ref{fig:xsect_decay} for a larger range of x-ray energies
than in Fig.~\ref{fig:xsection}.
The overall shape of our curve resembles the experimental result
in Fig.~1 of Schaphorst~\etal~\cite{Schaphorst:ME-93}.
However, our curve does not reach the same peak height of~$\approx
19 \kilobarns$ and it rises less steeply [our Fig.~\ref{fig:xsection}].
Considering the one-particle model adopted here, the agreement is satisfactory.
The photoabsorption cross section of hydrogen atoms above the $K$~edge is described
by a simple formula,
\eg,
Ref.~\onlinecite{Bethe:QM-57},
\begin{equation}
  \label{eq:xsect_decay}
  \sigma_{1s}(\omega_{\mathrm X}) = \sigma_{1s,\; \mathrm{a.e.}}
    \> \omega_{\mathrm X}^{-n} \; .
\end{equation}
For hydrogen, the exponent is~$n = 8/3 = 2.\bar6$ in the vicinity of the edge, \ie,
where the ejected electrons only have a small fraction of the $K$-shell energy.
It rises to the well-known~$n = 7/2 = 3.5$ far away from the edge for values
of~$\omega_{\mathrm X}$ that are 100~times or more the $K$~shell
energy~\cite{Bethe:QM-57}.
Formula~(\ref{eq:xsect_decay}) can be expected to well approximate the $K$-shell
photoabsorption cross section of more complex atoms like krypton~\cite{Bethe:QM-57}.
%
%
%
%
%
We obtain the constants~$\sigma_{1s,\; \mathrm{a.e.}} = 18623.8 \kilobarns \, \eV^n$
and $n = 2.63$ for the above-edge behavior of the cross section in
Eq.~(\ref{eq:xsect_decay}) by a nonlinear curve fit~\cite{More:LM-78} of the data
in Fig.~\ref{fig:xsect_decay}.
The exponent~$n$ is very close to the value derived for hydrogen, which
corroborates the assumption that also the above-edge behavior of the
Kr$\,1s$~cross section is well described by the theory and methods presented
in this paper.
Moreover, $n$~is consistent with a fit to the experimental attenuation cross
section of krypton in Ref.~\onlinecite{Saloman:XR-88} in the respective energy range.

\section{Conclusion}
\label{sec:conclusion}

We derive a formula for the x-ray photoabsorption cross section of an atom in the
light of a medium-intensity laser in the optical wavelength regime.
The laser and the x~rays are assumed to be linearly polarized.
The dressing laser effects a modification of the near-edge x-ray absorption and a
polarization dependence of the absorption cross section on the angle between the electric
field vectors of the individual light sources.

We use the Hartree-Fock-Slater approximation to describe the atomic
many-particle problem in conjunction with a nonrelativistic
quantum-electrodynamic approach to treat the light-electron interaction.
In order to deal with continuum electrons, a complex absorbing potential is employed that
is derived using the smooth exterior complex scaling technique.
Using the atomic orbitals and the number states of the free electromagnetic
fields in a product basis, a two-mode (or two-color) matrix representation
of the Hamiltonian is calculated.
A direct diagonalization of the complex symmetric matrix leads to the
transition rates for excitation out of the $K$~shell, which are
used to calculate the cross section.
Due to the relatively low intensity of existing third-generation synchrotron
sources, x-ray absorption may be described by a one-photon process.
This property is exploited in terms of a perturbative treatment to simplify
the two-mode problem, using both time-independent and time-dependent derivations
of the formula for the cross section.
The angular dependence of the cross section is studied and the approximate conservation
of its integral from~0 to $\infty$ is proven.

The theory is applied to a single krypton atom near the $K$~edge.
A pronounced modification of the energy-dependent cross section is
found with laser dressing.
The modification of the cross section depends notably on the angle
between the polarization vectors of laser and x~rays.
The behavior of the cross section above the edge is found to follow a well-known
approximation, thus reassuring the usefulness and quality of the theoretical framework.
Our theoretical predictions for noble-gas atoms (see also Ref.~\onlinecite{Buth:ET-up})
are presently under experimental investigation.

Our studies offer motivation and prospects for future research.
The theoretical framework of this article can easily be extended to multiphoton
x-ray processes for the emerging x-ray free electron lasers, \eg, the Linac Coherent
Light Source~(LCLS)~\cite{LCLS:CDR-02} at the Stanford Linear Accelerator Center~(SLAC).
Further investigations should treat the dependence of the photoabsorption
cross section on the wavelength and intensity of the laser.
Moreover, we used the electric dipole approximation in our derivations.
We expect nondipole effects to cause deviations from the angular dependence
of our formula for the total photoabsorption cross section.
This proposed route to study such effects is in fact much easier than the
conventional way to measure angular distributions of photoelectrons.
Furthermore, improvements of the description of the atomic many-particle problem
offer intriguing perspectives to study the interaction of light with
correlated electrons and the competition between the strength of both interactions.

\begin{acknowledgments}
We would like to thank Linda Young for fruitful discussions and a critical reading of
the manuscript.
C.B.'s research was funded by a Feodor Lynen Research Fellowship from the
Alexander von Humboldt Foundation.
R.S.'s work was supported by the Office of Basic Energy Sciences, Office of Science,
U.S.~Department of Energy, under Contract No.~DE-AC02-06CH11357.
\end{acknowledgments}

\end{document}